\documentclass[aps,prb,twocolumn,showpacs,groupedaddress]{revtex4}
\usepackage{graphicx}

\begin{document}


\title{Sagnac Rotational Phase Shifts in a Mesoscopic Electron Interferometer with Spin-Orbit Interactions}

\author{Marko Zivkovic, Markku J\"{a}\"{a}skel\"{a}inen, Christopher P. Search, and Ivana Djuric}
\affiliation{Department of Physics and Engineering Physics,Stevens Institute of Technology, Hoboken, NJ 07030}

\date{\today}

\begin{abstract}
The Sagnac effect is an important phase coherent effect in optical and atom
interferometers where rotations of the interferometer with respect to an inertial
reference frame result in a shift in the interference pattern proportional to the
rotation rate. Here we analyze for the first time the Sagnac effect in a mesoscopic
semiconductor electron interferometer. We include in our analysis Rashba spin-orbit
interactions in the ring. Our results indicate that spin-orbit interactions increase the
rotation induced phase shift. We discuss the potential experimental observability of the
Sagnac phase shift in such mesoscopic systems.
\end{abstract}

 \maketitle

\section{Introduction}
In the last decade experimental developments in mesoscopic condensed matter
and AMO (atomic, molecular, and optical) physics, such as the
explosive growth in semiconductor nanostructures, the creation of
atomic Bose-Einstein condensates (BEC) and ultra-cold atom
interferometers, and the interest in quantum computation and
information, have caused phase coherence and related phenomena to
receive extraordinary attention. Particularly interesting are
 quantum interference phenomena in ballistic
transport through high mobility nanostructures in which electron
propagation is described by quantum mechanics rather than by
classical transport. This has lead to novel experiments
with matter wave interferometers (MI's) for electrons
\cite{electron-interferometers} and quantum dot structures
\cite{quantum-dots} demonstrating quantum interference between
different paths.

Matter wave interferometry is a key paradigm for quantum interference and dates
back to the early electron-diffraction experiments. Recent advances show considerable promise for the development of new devices, mostly because the sensitivity of MI's
\cite{Clauser, Berman} far exceeds that of their optical counterparts for many important applications. Although both optical interferometers and MI's are able to detect rotations due
to the Sagnac effect, the sensitivity of atom-interferometer (AI)
based rotation sensors, however, can be as much as $Mc^{2}/\hbar \omega \sim 10^{10}$ times
greater \cite{Clauser, Scully} than that of optical ones \cite{Chow}. (Here $M$
is the atomic mass and $\hbar \omega$ is the energy of a photon.) Current
generation laboratory AI's \cite{Gustavson} already outperform
commercially available ring laser gyroscopes \cite{Chow}. Optical
gyroscopes are now used on virtually all commercial aircraft as well as on
spacecraft for inertial navigation. The potential improvement for rotation sensing with AI's, along with their ability to accurately detect small changes in gravitational fields, has resulted in intense
activity within the AMO community to develop AI sensors for inertial
navigation, geophysical prospecting, and tests of general relativity
\cite{Gustavson,McGuirk,Durfee}.

In 1913, Sagnac demonstrated that it is possible to detect rotations with
respect to an inertial frame of reference with an interferometer, using the
rotation-induced path length difference between its two arms.
The phase shift is easily understood if one considers a
ring shaped Mach-Zehnder interferometer of radius $R$ rotating about its axis
at the rate $\Omega$. In one arm of the interferometer, the particles are
co-propagating with the rotation, which increases the distance particles have
to travel before exiting by $\approx R\Omega t$. For the other arm, particles
are moving opposite to the direction of rotation and the distance they must
travel before exiting is decreased by the same amount. As a result, there is a
path length difference proportional to $\Omega$.

It should in principle be possible to observe this effect in another type of matter wave
device - electron interferometers (EI's). Mesoscopic semiconductor EI's have
been predominantly used for studying transport and quantum interference in low
dimensional systems \cite{electron-interferometers}. Recently there has been a
number of papers on their use to control and generate spin currents in the presence of spin-orbit(SO) coupling \cite{Junsaku,Konig, Frustaglia,shelykh,BranislavNikolic}. Surprisingly, there
has been no discussion of using them as gyroscopes. To date, the
only experiments on the rotation induced Sagnac effect for electrons were done with electron beams in
vacuum \cite{Hasselbach}. In comparison to optical or atom interferometers, EI's are much smaller, can be integrated with other solid
state devices, and are in many ways more robust due to the monolithic solid state structure.

The practical importance of the Sagnac effect for navigation combined with the technological advantages of solid state
devices, raises the question as to how easily this effect could be exploited in solid state EI's.
For electrons with effective mass $m^*\approx 0.1m_e$, the enhancement
factor relative to an optical interferometer with equal area is
$m^*c^2/\hbar\omega\sim 10^5-10^6$. On the other hand, the main disadvantage of
electron interferometers is the phase coherence length ${L_\Phi} \approx
l_{mfp}$, which  for electrons in solids limits the area of an interferometer to
approximately ${L_\Phi}^{2}/\pi$. Since the rotational phase shift is proportional to the enclosed area, this limitation implies a phase shift several orders of magnitude smaller than for current AI's \cite{Gustavson, Durfee}.  At the same time we note that each
order of magnitude improvement in the mean free path $l_{mfp}$, resulting from
improved fabrication techniques, yields a hundred-fold increase in the
maximum area, and as a consequence, in the rotational phase shift.
It is worth mentioning, however, that recently several papers have
pointed out that the rotation induced Sagnac effect could be observed in arrays of coupled
optical microring waveguides  by using 'slow' light \cite{peng,Scheuer}. The radii of the
microrings is $\sim 10 \mu m$, which is only about one
order of magnitude larger than already demonstrated semiconductor rings for
electrons and holes \cite{Konig,morpurgo,yau}. Recently, the Sagnac effect has been observed in the electronic conductance
of carbon nanotube loops with diameters $\sim 1\mu m$  although the origin of the Sagnac phase difference was not
due to an externally applied rotation of the loops \cite{refael}.

The main goal of this paper is to investigate a way to enhance the Sagnac phase shift to readily
detectable values. To this end, we analyze the coherent interplay of the Sagnac effect and Rashba
spin-orbit interaction and estimate the resulting enhancement of the Sagnac phase shift. Indeed, we find that the
interplay between the spin interference driven by the spin-orbit interaction and the Sagnac effect result in a larger
phase shift for a given rotation rate. This increase in the phase shift can be interpreted as a larger effective area for the interferometer.

The paper is organized as follows: Section \ref{Model} establishes the model and
introduces the slowly varying envelope approximation as a mathematical technique for
solving the Schr{\"o}dinger equation in the ring. To justify the applicability of the
SVE, we compare our results to exact numerical solutions of the Schr{\"o}dinger equation
for several parameter values. In Sec. \ref{Results} we present the results of our
simulations, and calculate the enhancement of rotational phase shifts.  We also discuss
the effect of quantum noise on the detectability of rotational phase shifts. Finally,
Sec. \ref{Discussion} is a summary and outlook where we discuss how to optimize the phase
shift by integrating a series of EI's into an array.

\section{Theoretical Model}\label{Model}
We consider a quasi-one-dimensional ring of radius $r_0$, which could be defined in a
two-dimensional electron \cite{Konig, morpurgo} or hole \cite{yau} semiconductor
heterostructure [Fig. 1(a)]. We presume that the arms of the ring behave as a ballistic
conductor (i.e. the length of the arms is smaller then the electron mean free path). The
ring is coupled to two electron reservoirs with a bias voltage $V_1-V_2$ resulting in a
current $I=G(V_1-V_2)$. In the growth direction (z-axis), which is perpendicular to the
plane of the ring, a static magnetic field ${\bf B}=\nabla \times {\bf A}$ and electric
field ${\bf E}$ are applied. The electric field comes from the electrostatic potential of
a biased gate above the plane of the ring and has no contribution from the static vector
potential ${\bf A}$. Due to the applied magnetic field {\bf B}, there is a nonzero Zeeman
splitting between electron spin states as well as a finite magnetic flux through the ring
that would give rise to Aharonov-Bohm oscillations.

In semiconductor heterostructures with structure inversion asymmetry, such as
InGaAs/InAlAs \cite{Nitta2} or  HgTe/HgCdTe \cite{Konig} quantum wells, the
dominant spin-orbit interaction is given by the Rashba Hamiltonian
\cite{Rashba},
\begin{equation}
H_{int}=\alpha {\bf{\sigma}} \cdot {\bf E} \times \Pi =\alpha_R\hat{z}\cdot({\bf \sigma} \times {\Pi})
\label{Hrashba}
\end{equation}
where $\sigma$ is the vector of the Pauli spin operators,  $\Pi={\bf p}-e{\bf
A}$ the electron momentum, and ${\alpha_R}=\alpha E_z$ the Rashba constant. For
electrons traveling around the ring, {\bf E} gives rise to a momentum
dependent magnetic field ${\bf B}_R$ in the plane of the ring due to the SO
coupling of the electron spin with its center-of-mass motion. An important
feature of the Rashba interaction is that the strength of the SO interaction is
proportional to the external electric field, which enables easy control by the
gate above the ring. The spins precess around ${\bf
B}_{eff}={\bf B}+{\bf B}_R$ [Fig. 1(b)] as they propagate around the ring. This
leads to interference between the spin directions of an electron whose wave
function is coherently split between the two paths of the interferometer and
then later coherently recombined upon exiting. Note that because we consider only ballistic
transport here, the Rashba term only gives rise to coherent coupling between the spin states and
does not cause dephasing of the spin coherence due to scattering of the orbital wave function.

\begin{figure}[htb]
\centering
\includegraphics[width=0.4\textwidth]{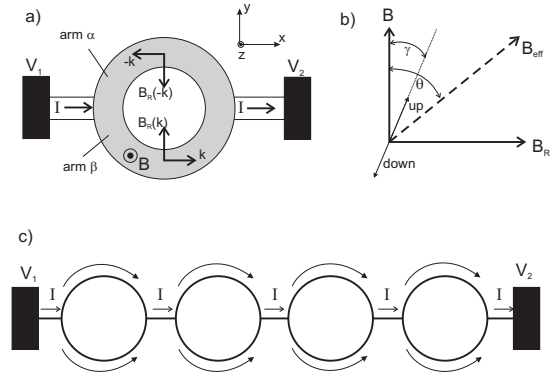}
\caption{(a) Schematic diagram of an electron interferometer: 1D ring of radius
$r_0$ subject to Rashba spin-orbit coupling and in the presence of an external
magnetic field ${\bf B}$. (b) Effective magnetic, ${\bf B}_{eff}={\bf B}+{\bf
B}_r$ field that spins perceive while traveling around the ring. (c) A one
dimensional array of ring interferometers in series. }\label{PIC.1}
\end{figure}

The effective 1D Hamiltonian for electrons (charge $e<0$ and effective mass $m^*$) propagating in ring subject to Zeeman
and Rashba coupling, with coupling constants $\mu$ and $\alpha_{R}$, respectively, is
\cite{Frustaglia,Meijer},
\begin{eqnarray}
\hat{H}_{1D}(\varphi)=\frac{\hbar \omega_0}{2} (-i \frac{\partial}{\partial \varphi} + \frac{\Phi}{\Phi_0})^2
+\frac{\hbar \omega_B}{2}\sigma_Z
                        \nonumber\\
+ \frac{\hbar \omega_R}{2}(\cos\varphi \sigma_x +\sin\varphi \sigma_y)(-i\frac{\partial}
{\partial \varphi} + \frac{\Phi}{\Phi_0})
                        \nonumber\\
- i\frac{\hbar \omega_R}{4}(\cos\varphi \sigma_y - \sin\varphi \sigma_x)
\label{H1D}
\end{eqnarray}
where the frequencies $\omega_0=\hbar/(m^*r_0^2)$, $\omega_B=2 \mu B/\hbar$, and $\omega_R=2\alpha_R/\hbar r_0$,
$\Phi=\pi r_0^2 B$, and the flux quantum $\Phi_0=h/e$ have been introduced. Here we have made use of the form of the vector potential for a
uniform B-field in the z-direction, ${\bf A}={\bf \hat{e}}_\varphi rB/2$, to reexpress all quantities involving ${\bf A}$ in terms of the magnetic flux through the ring.

If the ring is rotating with angular velocity $\Omega$ about the axis perpendicular to
the ring, the effective distance that particles have to travel before exiting the ring is
increased by $\delta l_{\alpha}={r_0}\Omega t_{\alpha}$ for particles co-propagating with
the rotation and decreased by the amount $\delta l_{\beta}={r_0}\Omega t_{\beta}$ for
particles that are moving in the opposite direction. Here we assume that particles going
from left to right in the upper arm ($\alpha$) in Fig. 1(a) are co-propagating with the
rotation while those going in the same direction through the lower arm ($\beta$) are
counter-propagating. For small $\Omega$ such that $\delta l_{\alpha (\beta)} \ll
l_{\alpha (\beta)}$ where $l_{\alpha (\beta)}$ is the path length of the upper
co-propagating (lower counter-propagating) arm, then $t_{\alpha (\beta)}=(l_{\alpha
(\beta)}\pm \delta l_{\alpha (\beta)} )/v \approx \l_{\alpha (\beta)}/v= \pi {r_0}/v$
where $v$ is the velocity of the particles. This causes a Sagnac phase difference between
two counter-propagating de Broglie waves in the ring of $\Delta\phi= k[(l_{\alpha}+\delta
l_{\alpha}) -(l_{\beta}-\delta l_{\beta})]=k 2 \pi {{r_0}^2} \Omega /v =2A \Omega
m^*/\hbar$, where $k=m^{*}v/\hbar$ is the wave number of an electron, $A={{r_0}^2}\pi$ is
the area enclosed by the arms of the interferometer\cite{Scully}. This derivation of the
Sagnac phase shift assumes that the spin of the particle is not affected by the rotation.
However, in addition to the normal Sagnac phase shift, the rotation of the ring changes
the distance that the spins precess around ${\bf B}_{eff}$ as they propagate along the
two arms. The relative orientations of the spins from the two arms when recombined has
now been changed as a result of the rotation. The resulting spin interference will give a
contribution to the Sagnac phase shift, $\Delta\phi$, that is a function of $\Omega$ and
$|{\bf B}_{eff}|$.

When the ring is rotating, the system could be described by the same
Hamiltonian as the one given in Eq. (\ref{H1D}), but the point where the two
counter-propagating electron waves recombine and interfere would change its
position with time. An easier way to analyze interference in a rotating ring is
to change the reference system in which we observe the process from the
non-rotating to the rotating one. In the rotating frame of reference, the
angular momentum of particles co-propagating with the rotation is decreased
while those that are counter-propagating  is increased, similar to the Doppler
effect. The wave functions in the two reference frames are related by
$\Psi_R=\hat{R}\Psi$, where $\Psi_R$ and $\Psi$ are the wave functions in the
rotating and non-rotating frames, respectively, and $\hat{R}=\exp[i\Omega t
\hat{n} \cdot \vec{L}/\hbar]$ is the rotation operator ($\hat{n}$ is the rotation
axis and $\vec{L}$ the angular momentum operator). Only rotations around the axis perpendicular to the plane
of the ring will result in a relative phase shift between the two arms. For this reason we set $\vec{L}\rightarrow L_z$ without loss of generality.
The Hamiltonian for an electron in the rotated frame is then given by:
\begin{equation}
\hat{H}_R (\varphi)= \hat{H}_{1D} (\varphi) + i\hbar\Omega\frac{\partial}{\partial \varphi}
\label{H_R}
\end{equation}

The energy eigenfunctions can be expressed in the following form:
\begin{equation}
\Psi_R(\varphi,t)=e^{-\frac{iE}{\hbar}t}\Psi_R(\varphi)=e^{-\frac{iE}{\hbar}t} \left[ \begin{array}
{ccc} S_\uparrow(\varphi) \\
S_\downarrow(\varphi) \end{array} \right ] e^{iKr_0\varphi} \label{ansatz}
\end{equation}
where $S_\uparrow, S_\downarrow$ are the angular dependent spinors for spin states
oriented along the z-axis with energy $E$ and momentum $K$ propagating inside the ring
with radius $r_0$. This is inserted into the time-dependent Schr{\"o}dinger equation for
the Hamiltonian in Eq. (\ref{H_R}), giving us a system of second order differential
equations for the envelope function $S_\uparrow, S_\downarrow$. If the envelopes
functions are smooth functions that vary much slower than the carrier wave,
\[
|\partial S_{\sigma}(\varphi)/ \partial \varphi| \ll Kr_0
|S_{\sigma}(\varphi)|,
\]
we can neglect the second order derivatives. This is known as the slowly
varying envelope (SVE) approximation in optics \cite{Meystre}. While SVE is a widely used
technique in nonlinear and atom optics, it is not common in mesoscopic
transport. With this approximation the system becomes,
\begin{widetext}
\begin{equation}
\left [ \begin{array}{ccc} \dot{S_\uparrow} \\ \dot{S_\downarrow} \end{array} \right ] = M \left [ \begin{array}
{ccc} i \left ( aP_1 - \left(\frac{\omega_R}{\omega_0}\right)^2 (b+1) \right ) & i(ab-P_2)\frac{\omega_R}{\omega_0}
e^{-i\varphi} \\
i(a(b+1)-P_1)\frac{\omega_R}{\omega_0}e^{i\varphi} & i\left (aP_2-\left(\frac{\omega_R}{\omega_0}\right)^2 b \right)
\end{array} \right ] \left[ \begin{array}{ccc} S_\uparrow \\ S_\downarrow \end{array} \right]
\label{system}
\end{equation}
\end{widetext}
where dots over $S_\uparrow, S_\downarrow$ denote derivatives with respect to
the angular position in the ring $\varphi$, and
\begin{center}
\begin{eqnarray}
M =  \left ( \left(\frac{\omega_R}{\omega_0} \right)^2 -a^2 \right)^{-1},\qquad \quad \nonumber\\
a=2\left(Kr_0+\frac{\Phi}{\Phi_0}-\frac{\omega}
{\omega_0}\right), b=Kr_0+\frac{\Phi}{\Phi_0}-\frac{1}{2}, \nonumber \\
P_{1/2}=\left[2\frac{2E}{\hbar \omega_0}+\left(Kr_0+\frac{\Phi}{\Phi_0} \right)^2 - 2Kr_0\frac{\omega}{\omega_0} \pm
\frac{\omega_B}{\omega_0} \right] \nonumber
\label{Mab}
\end{eqnarray}
\end{center}
These coupled first order differential equations are numerically easier to integrate than
the second order coupled equations for $S_{\sigma}$ that would be directly obtained from
the Schrodinger equation.

In the Landauer-Buttiker formalism, the zero-temperature conductance of a
ballistic conductor is given by:
\begin{equation}
G=\sum_{\sigma',\sigma} G^{\sigma'\sigma}=\frac{e^{2}}{h} \sum_{\sigma',\sigma} \sum_{m',m=1}^M T_{m'm}^{\sigma'\sigma}
\label{conductance}
\end{equation}
where $T_{m'm}^{\sigma'\sigma}$ denotes the quantum mechanical probability of
transmission between incoming $(m,\sigma)$ and outgoing $(m',\sigma')$ asymptotic states.
The labels $m, m'$ and $\sigma, \sigma'$ refer to the corresponding orbital mode and spin
quantum numbers, respectively. From Eq. (\ref{conductance}) it can be seen how a change
of the transmission coefficients due to interference from rotation induced phase shifts
causes a modulation of the current through the ring.  For convenience, we will restrict
our discussion to a single orbital mode and drop the subscripts for the transmission
probabilities.

By specifying the spin states of the electrons when they enter the ring, $S_{\sigma,
\alpha(\beta)}(0)$, we can obtain $S_{\sigma', \alpha(\beta)}(\pm \pi)$ at the end points
of the interferometer arm where the wave function is recombined. From this the
transmission coefficients, $T^{\sigma'\sigma}$, and hence the conductance can be directly
calculated. For example, if spin up polarized current enters the ring and the wave
functions is equally split between the two arms,
$S_{\uparrow,\alpha}(0)=S_{\uparrow,\beta}(0)=1/\sqrt{2}$, then the probability of
measuring a spin down electron leaving the ring on the other side is then
$T^{\downarrow,\uparrow}=|S_{\downarrow,\alpha}(\pi)+S_{\downarrow,\beta}(-\pi)|^2/4$.

If the leads connected to the ring are unpolarized, i.e. the leads are an incoherent
mixture of spin-up and down, then it is only the total charge conductance that will be
measured. On the other hand, the field of spintronics has been making rapid progress
towards methods for generating and measuring spin polarized currents by such methods as
ferromagnetic leads and the spin hall effect \cite{zutic,awschalom,tinkham}. One can then
imagine that incident on the ring from the left lead is a current that is spin polarized
along the z-direction and that in the second lead one can measure the spin polarization
of the current exiting the ring. In this case, one is directly measuring the spin
polarized conductances, $G^{\sigma'\sigma}$. In the next section we consider both
scenarios.

\begin{figure}[htb]
\centering
\includegraphics[width=0.48\textwidth]{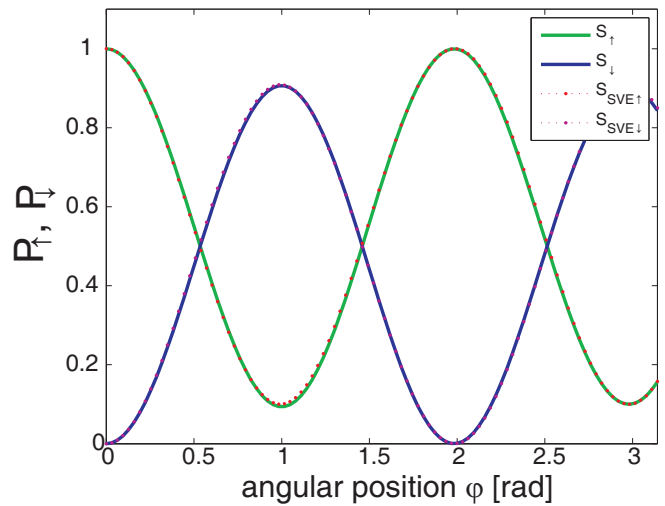}
\caption{Comparison between SVE and numerical solution of Schrodinger equation
without SVE: Probabilities for spin up ($P_{\uparrow}=|S_{\uparrow}(\varphi)|^2$) and spin down ($P_{\downarrow}=|S_{\downarrow}(\varphi)|^2$)
states, along one arm of
interferometer assuming spin-down electrons enter that arm. Here
$r_0=1000nm$, $Q_R=3$, and there is no external magnetic field. }\label{PIC.2}
\end{figure}

\section{Results}\label{Results}
In our simulation of the electron interferometer we used $r_0=1000nm$ for the radius of
ring and for the electron we chose an effective mass $m^*=0.067 m_0$ and wave number
$K=0.1nm^{-1}$. In addition to this we focus on ${\bf B}=0$ from here on since this is
expected to produce the maximal spin interference between the two arms. We solved Eq.
(\ref{system}) numerically using the SVE approximation, and in order to check its
validity we did the same calculation including the second order derivatives. The
comparison between the two methods is shown in Fig.\ref{PIC.2}, where we see that the
difference between results derived without the approximation (solid line) and with the
SVE approximation (dashed line) is negligibly small. The SVE approximation is justified
only when $1/K$ is much less than the distance over which the envelope functions change
significantly, which is given by the spin precession length, $\ell_{SO}=\hbar\pi/\alpha_R
m^*$. In terms of $\alpha_R$ and $K$ this condition is then,
\[
\frac{\hbar\pi K}{\alpha_R m^*} \gg 1
\]
Application of this mathematical technique in the future can considerably reduce
calculation time needed for more complex problems. In particular, it provides an
alternative to other techniques currently used such as numerical evaluation of real space
Greens functions \cite{BranislavNikolic}.

The spatial Rabi oscillations between spin states in Fig. 2 are not of full amplitude.
This is because the diagonal terms in Eq. \ref{system} are not the same, which means that
the slowly varying envelope functions are not degenerate. It is well known from the
theory of two-level quantum systems that the amplitude of the oscillations between states
decreases with increasing energy difference between the states. How this 'energy
difference' arises from the underlying Hamiltonian Eq. \ref{H1D} can be understood by
noting that to obtain Eq. \ref{system} for the stationary states in an arm of the ring,
we have expressed the spatial equations of motion in the form
\[
i\hbar\frac{\partial}{\partial \varphi} \Psi=\tilde{H}\Psi.
\]
This has the form of a Scrodinger equation but with the substitution $t\rightarrow \varphi$. To obtain this form from Eq. \ref{H1D} one must
multiply both sides by $[\frac{\hbar \omega_R}{2}(\cos\varphi \sigma_x +\sin\varphi \sigma_y)]^{-1}$, which gives rise to terms proportional to $\sigma_z$.

\begin{figure}[htb]
\centering
\includegraphics[width=0.5\textwidth]{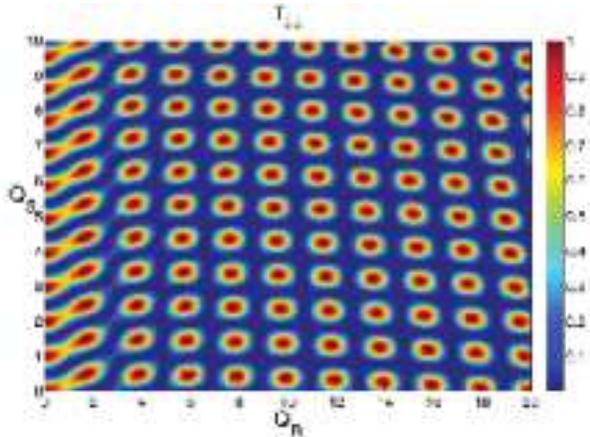}
\caption{Spin polarized transmission coefficient, $T^{\downarrow,\downarrow}$, for an
electron propagating through EI for $B=0$, for different values of Rashba SO coupling
strength ($Q_R=\omega_R/\omega_0$) and Sagnac strength ($Q_S=\Omega/\omega_0$). This is
the probability that an electron with spin down at the entrance of the ring has the same
spin when it exits the ring} \label{PIC.3}
\end{figure}

\begin{figure}[htb] \centering
\includegraphics[width=0.5\textwidth]{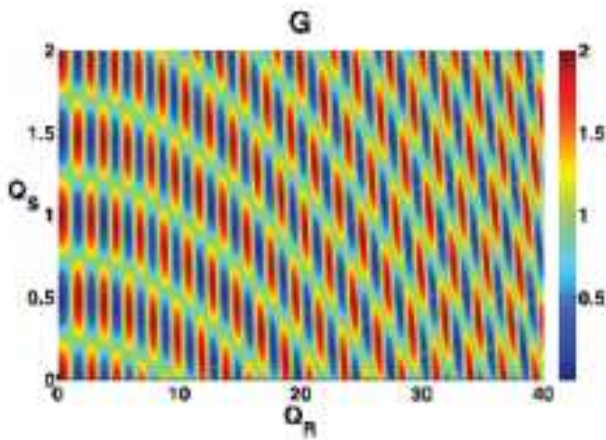}
\caption{Total conductance in units of $e^2/h$ for an electron propagating through EI for
$B=0$ and different values of Rashba SO coupling strength ($Q_R=\omega_R/\omega_0$) and
Sagnac strength ($Q_S=\Omega/\omega_0$).} \label{PIC.3}
\end{figure}

Having found values for $S_\uparrow$ and  $S_\downarrow$ we determined the transmission
coefficients for different magnitudes of Rashba SO coupling and rotation rates. Figures 3
and 4 shows $G^{\downarrow,\downarrow}$, assuming a spin down polarized current incident
on the ring, as well as the total conductance, $G$, for unpolarized currents in units of
$e^2/h$, respectively. One can conclude from both figures that the Rashba and Sagnac
effect do not give rise to separate contributions to the transmission phase since the
interference pattern does not lie along horizontal or vertical lines.

Let us focus on Fig. 3, which involves only a single transmission probability that can be
written in the form $T^{\sigma,\sigma}\propto \cos^2 (\Delta\phi)$. Now let us assume
that the phase shift can be written as $\Delta\phi=f(Q_S)+g(Q_R)$ where $f(Q_S)$ and
$g(Q_R)$ are some functions of the dimensionless rotations rate, $Q_S=\Omega/\omega_0$,
and the dimensionless Rashba term $Q_R=\omega_R/\omega_0$. In this case one can see that
if $Q_R (Q_S)$ is fixed and $Q_S (Q_R)$ allowed to vary, which corresponds to moving
along a vertical (horizontal) line in Fig. 3, then the minima and maxima of the
interference pattern should lie entirely along the vertical (horizontal) line. As one can
see, however, the minima and maxima of the interference pattern follow lines that deviate
from vertical and horizontal. This indicates that the phase of $T^{\sigma,\sigma}$ is a
nonlinear combination of $Q_R$ and $Q_S$.

\begin{figure}[htb]
\centering
\includegraphics[width=0.48\textwidth]{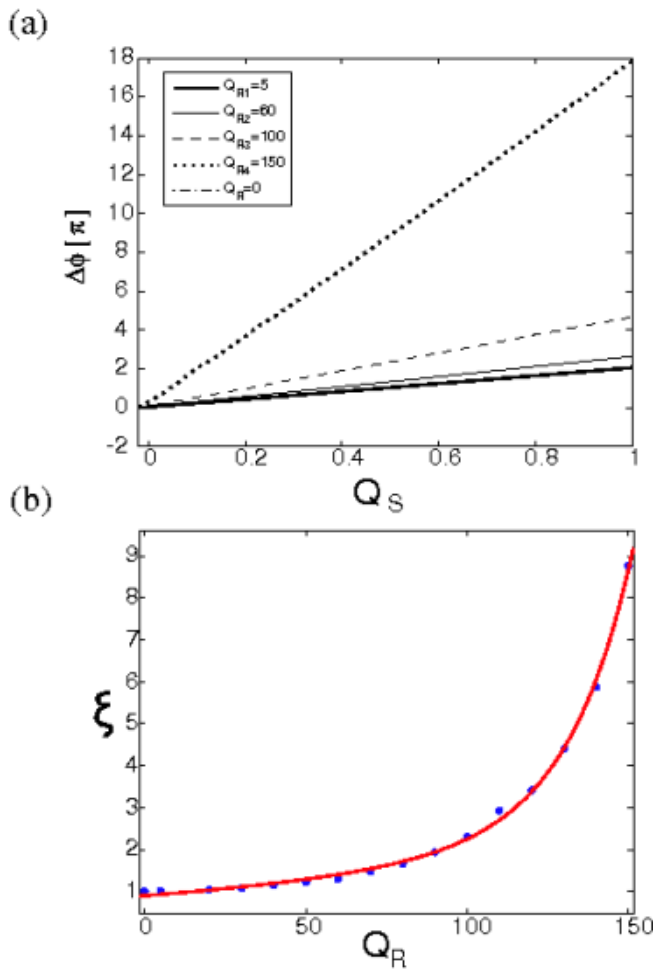}
\caption{ (a) Sagnac rotational phase shift as a function of the rotation rate
$Q_S=\Omega/\omega_0$ for different values of the Rashba SO interaction strength, $Q_R$.
Phase shift, $\Delta \phi$ is in units of $\pi$ radians (b) Dimensionless enhancement
factor, $\xi$, as a function of $Q_R$. Dots are slopes of the curves in (a) as well as
for other values of $Q_R$ not shown in (a). The solid line is a numerical fit to  the
points.}\label{PIC.4}
\end{figure}

\begin{figure}[htb]
\centering
\includegraphics[width=0.48\textwidth]{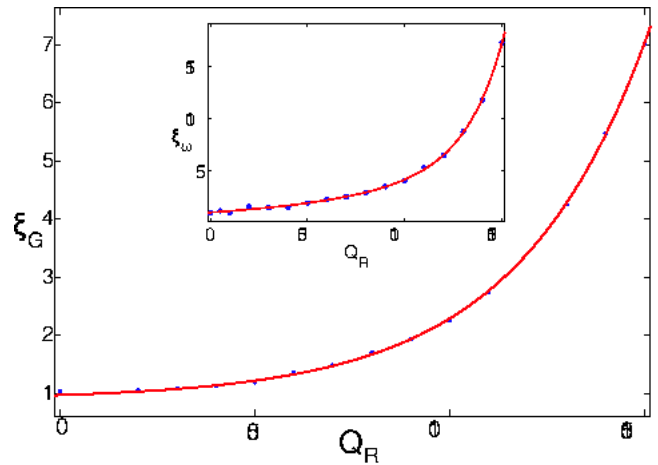}
\caption{ Dimensionless enhancement factor for the total conductance,
$\xi_G=\left[\partial G(Q_R,Q_S)/\partial Q_S\right]_{max}/\left[\partial
G(Q_R=0,Q_S)/\partial Q_S\right]_{max}$, as a function of $Q_R$. Dots are numerically
calculated values of $\xi_G$ for different $Q_R$ The solid line is a numerical fit to the
points. Inset, $\xi_{\omega}(Q_R)=T(Q_R=0)/T(Q_R)$ is the frequency of the oscillations
in $G$ as a function of $Q_S$ for different $Q_R$. Here, $T(Q_R)$ is the period of the
oscillations in $G$ for fixed $Q_R$ as $Q_S$ is increased.}\label{PIC.4}
\end{figure}

In the case of spin polarized transport, the rotational phase shift $\Delta\phi$ can be
uniquely equated with the phase of the interference pattern in Fig. 3 since it results
from only a single transmission probability. In Fig. 5 we have extracted $\Delta\phi$ as
a function of $\Omega$ for different values of $Q_R$ from our numerical results for
$T^{\downarrow,\downarrow}$. The Sagnac phase shift with no Rashba effect is given by the
dash-dotted line, which is indistinguishable from the line for $Q_R\!\!=\!\!5$ (thick
solid line). This confirms that for weak SO coupling ($Q_R\lesssim 10$) there is only
negligible mixing between the SO coupling and the rotational phase shift. For higher
values of $Q_R$, the mixing becomes stronger, which is manifested by a steeper slope. Our
numerical results indicate that the rotation induced phase shift is approximately
\begin{equation}
\Delta\phi\cong \xi 2\pi r_0^2 \Omega m^*/\hbar \label{polarized-enhancement}
\end{equation}
where $\xi \geq 1$ is an enhancement factor due to the SO coupling, which is shown in
Fig. 5(b), and for which a numerical fit yields,
\[
\xi(Q_R)\approx 0.9\exp(0.007Q_R)+0.003\exp(0.05Q_R)
\]
By increasing $Q_R$, it is possible to more easily detect small changes in the angular
velocity.

By contrast, the total conductance involves a summation of four transmission
probabilities that do not necessarily oscillate in phase with each other. In this case it
is harder to define the rotation induced phase shift. However, the quantity that is of
most interest experimentally is how much the conductance changes due to a small change in
$Q_S$, $\Delta G\approx \left(\partial G/ \partial Q_S\right)\Delta Q_S$. This allows us
to define an enhancement, $\xi_G$,
\begin{equation}
\xi_G(Q_R)=\frac{\left[ \partial G(Q_R,Q_S)/ \partial Q_S\right]_{max}}{\left[ \partial
G(Q_R=0,Q_S)/ \partial Q_S\right]_{max}} \label{G-enhancement}
\end{equation}
where $max$ means the maximum magnitude of the slope for fixed $Q_R$. It is worth noting
that if we assume a simple interference pattern of the form $G=A\cos^2[\Delta\phi]$ and
use $\Delta\phi$ in Eq. \ref{polarized-enhancement}, then we obtain from Eq.
\ref{G-enhancement} $\xi_G=\xi$, which shows that Eq. \ref{G-enhancement} is consistent
with our definition of $\xi$ for the spin polarized case. Fig. 6 shows $\xi_G$ as a
function of $Q_R$. As one can see, $\xi_G\sim \xi$ for all $Q_R$. Thus the enhancement
can just as easily be seen in the total conductance. Finally, the inset of Fig. 6 shows
the oscillation frequency for the interference pattern in $G$ for different values of
$Q_R$ (moving along vertical lines in Fig. 4). As one can see, the oscillation frequency
increases more rapidly than $\xi_G$. This is because the amplitude of the oscillations
decrease at a rate that is smaller than the rate of increase in the frequency. As a
result $\xi_G$ increases but more slowly than the oscillation frequency.

The enhancement factors, $\xi$ and $\xi_G$, are a result of the different spin
orientations of electrons created in the two arms. The spin of electrons going through
the upper arm precess around ${\bf B}_{eff}$ by a larger angle before exiting as compared
to the lower arm. This is due to the longer path length of the upper arm. As a result,
the orientations of the spins from the upper and lower arms are different when recombined
at the second lead and this imbalance in the spin precession angles changes the spin
resolved conductances. Recently it was demonstrated \cite{Gvozdic} that by using holes
instead of electrons it is possible to increase the strength of the Rashba interaction by
about three orders of magnitude. Based on the results presented here, such extremely
large Rashba strengths ($Q_R\sim 1000-10000$) should lead to Sagnac phase shifts that are
orders of magnitude larger than shown here. However, for such large $Q_R$ the SVE
approximation breaks down and new numerical techniques must be sought.

The minimum detectable phase difference in matter wave interferometers, $\Delta
\varphi_{min}$, is determined by the quantum fluctuations in the
measured phase difference. These fluctuations are the result of the partition
noise (also referred to as shot noise) that results from the splitting and
recombining of the particles at the beam splitters. For uncorrelated particles
the noise is Poissonian and the minimum detectable phase shift is \cite{Scully}
\begin{equation}
|\Delta\varphi_{min}| = \frac{1}{\sqrt{N}} \label{phasedifference}
\end{equation}
where $N$ is the total number of particles that pass through the interferometer during
the measurement time. This result ignores quantum statistics. If quantum statistics are
accounted for, it is found that $|\Delta\varphi_{min}|$ continues to scale like
$N^{-1/2}$ for bosons and fermions \cite{search-meystre}. The number of electrons passing
through the ring per unit time is proportional to the current through the ring $I$. By
equating the rotational phase shift with the shot-noise limited minimum detectable shift,
$|\Delta \phi|=|\Delta\varphi_{min}|$ and using Eqs. (\ref{phasedifference}) and
(\ref{polarized-enhancement}), we find that the minimum detectable rotation rate
$\Omega_{min}$ is
\begin{equation}
\Omega_{min}\cong \frac{\hbar}{\xi 2\pi r_0^2 m^*} \left (\frac{I t_m}{|e|}
\right )^{-\frac{1}{2}} \label{omegamin}
\end{equation}
where $t_m$ is the measurement time. Strong SO interaction yields $\xi \gg 1$ and reduces
$\Omega_{min}$ accordingly. However, even if we take $\xi=1$ corresponding to no SO
interactions and a ring of radius $10\mu m$ with $I=100nA$, one finds that
$\Omega_{min}=3.48 t_m^{-1/2} rad/s$ with $t_m$ measured in seconds. The quantum shot
noise represents the only fundamental physical limit to the phase resolution. However,
even if the ring itself is limited only by shot noise, the electrical current from the
ring must be amplified to more readily detectable values. Current cold amplifiers have
noise that is still well above the shot noise limit although recent experiments have
demonstrated novel very low noise mesoscopic amplifiers based on single electron
transistors \cite{wu} and Josephson junctions \cite{delahaye}. Also, theoretical work has
shown how to reach the quantum limit in linear electronic amplifiers \cite{gavish}. It is
then reasonable to assume then that future generation amplifiers will reach the quantum
noise limit.

\section{Conclusion}\label{Discussion}

This is the first study of the Sagnac effect in solid state electron ring conductors. We
have demonstrated that the SVE approximation is justified for typical spin-orbit coupling
strengths and also shown that the Rashba spin-orbit interaction can enhance the
sensitivity of rotation measurements. The spin-orbit enhancement can be regarded as an
increased effective area for the interferometer. Moreover, our estimates indicate that
the Sagnac phase shift can easily be made larger than the quantum shot noise limit, which
is the only fundamental obstacle. It is our hope that this work will stimulate further
interest in this problem and that next generation experiments will be able to measure the
Sagnac effect in semiconductors.

Another possible method for enhancing the Sagnac effect is to use a serial array of $N$
ring interferometers as depicted in Fig. 1(c). Transport within each ring is assumed to
be ballistic. In this case the resistivity of the rings (ignoring the contact resistance
and the resistance of the channels connecting the rings) is given by (ignoring for the
sake of simplicity here spin dependent transport)\cite{datta}
\begin{equation}
G_{rings}^{-1}=\frac{h}{2e^2}\sum_{i=1}^{N}\frac{1-T_i}{T_i} =\frac{h}{2e^2}\sum_{i=1}^{N}\tan^2 \left(\Delta\phi^{(i)}/2 \right)
\label{Gmin1}
\end{equation}
where $T_i=\cos^2(\Delta\phi^{(i)}/2)$ is the transmission probability through the
$i^{th}$ ring with Sagnac phase shift $\Delta\phi^{(i)}$. For small phase shifts and
ignoring differences between the rings, one sees that the resistance is
$G_{rings}^{-1}\propto N(\Delta\phi /2)^2$. If we do not assume ballistic transport
between the rings, this device should be scalable to large $N$ since then the total size
of the array can be $\gg l_{mfp}$.  Even though the phase shift in each ring may be too
small too measure, the effect is compounded as the electron passes through each
successive ring resulting in a phase shift that is enhanced by $\sqrt{N}$ in comparison
to that of a single ring. A similar idea was proposed for light propagating coherently in
a two dimensional array of coupled microring optical waveguides \cite{Scheuer} where the
enhancement relative to a single ring was found to be $N^2$. One of our goals in a future
publication is to explicitly calculate the contribution to the resistance due to the
channels connecting the rings assuming either incoherent transport or ballistic transport
between rings as well as the effect of SO coupling in the rings.

\end{document}